\documentclass[sigconf]{acmart}

\AtBeginDocument{%
  \providecommand\BibTeX{{%
    \normalfont B\kern-0.5em{\scshape i\kern-0.25em b}\kern-0.8em\TeX}}}

\setcopyright{acmcopyright}
\copyrightyear{2023}
\acmYear{2023}
\acmDOI{XXXXXXX.XXXXXXX}

\acmConference[MM ’23]{}{October 29-- November 02, 2023}{Ottawa, Canada}

\acmPrice{15.00}
\acmISBN{978-1-4503-XXXX-X/18/06}
\usepackage{multirow}
\usepackage{algorithm} 
\usepackage{algpseudocode}
\usepackage{subcaption}
\usepackage{balance} 
\usepackage[shortlabels]{enumitem}
\copyrightyear{2023}
\acmYear{2023}
\setcopyright{acmlicensed}\acmConference[MM '23]{Proceedings of the 31st
ACM International Conference on Multimedia}{October 29-November 3,
2023}{Ottawa, ON, Canada}
\acmBooktitle{Proceedings of the 31st ACM International Conference on
Multimedia (MM '23), October 29-November 3, 2023, Ottawa, ON, Canada}
\acmPrice{15.00}
\acmDOI{10.1145/3581783.3612443}
\acmISBN{979-8-4007-0108-5/23/10}
\begin{document}

\title{Predictive Sampling for Efficient Pairwise Subjective Image Quality Assessment}
\author{Shima Mohammadi}

\affiliation{%
  \institution{Instituto Superior Técnico - Instituto de Telecomunicações}
  \city{Lisbon}
  \country{Portugal}
}\email{shima.mohammadi@lx.it.pt}

\author{Jo\~{a}o Ascenso}
\affiliation{%
  \institution{Instituto Superior Técnico - Instituto de Telecomunicações}
  \city{Lisbon}
  \country{Portugal}}
\email{joao.ascenso@lx.it.pt}

\begin{abstract}
Subjective image quality assessment studies are used in many scenarios, such as the evaluation of compression, super-resolution, and denoising solutions. Among the available subjective test methodologies, pair comparison is attracting popularity due to its simplicity, reliability, and robustness to changes in the test conditions, e.g. display resolutions. The main problem that impairs its wide acceptance is that the number of pairs to compare by subjects grows quadratically with the number of stimuli that must be considered. Usually, the paired comparison data obtained is fed into an aggregation model to obtain a final score for each degraded image and thus, not every comparison contributes equally to the final quality score. In the past years, several solutions that sample pairs (from all possible combinations) have been proposed, from random sampling to active sampling based on the past subjects' decisions. This paper introduces a novel sampling solution called \textbf{P}redictive \textbf{S}ampling for \textbf{P}airwise \textbf{C}omparison (PS-PC) which exploits the characteristics of the input data to make a prediction of which pairs should be evaluated by subjects. The proposed solution exploits popular machine learning techniques to select the most informative pairs for subjects to evaluate, while for the other remaining pairs, it predicts the subjects' preferences. The experimental results show that PS-PC is the best choice among the available sampling algorithms with higher performance for the same number of pairs. Moreover, since the choice of the pairs is done \emph{a priori} before the subjective test starts, the algorithm is not required to run during the test and thus much more simple to deploy in online crowdsourcing subjective tests.
\end{abstract}



\maketitle
\vspace{-8pt}
\section{Introduction} \label{section-Introduction}
Subjective assessment methodologies and studies are essential for evaluating the visual quality obtained with different image processing or computer vision algorithms. All subjective studies rely on human subjects to provide their perception of image quality and provide insights about the visual experience of an image, and measure different image quality aspects, e.g., in the evaluation of image compression techniques. Nowadays, there are several subjective test methodologies \cite{testolina2021review} recognized as reliable but most often a single or double stimulus methodology is used. In such cases, a panel of human subjects is required to attribute scores to each visual stimulus in a predefined category scale, and the average scores of all the subjects are reported as mean opinion scores (MOS) or differential MOS with respect to a predefined reference.

However, the interpretation of the category scales might be different among the subjects, or the subjects might change their decision during the test, yielding inconsistent labels. For example, subjects may have different opinions about an "\textit{Excellent}" stimulus, or given another previous stimulus, they might change their opinion. Later on, the collected scores are used as ground-truth to benchmark some objective assessment models, or to train a learning-based objective metric and thus, accurate and reliable scores are essential. Absolute category scale (ACR) or double stimulus impairment scale subjective tests fall in this category. Despite their popularity, they have some disadvantages such as lack of sensitivity: limited in their ability to detect small differences in quality, and response bias: scores may be influenced by the subjects' personal biases, past experiences, or expectations, which may lead to inaccurate results and thus limits the reliability of the tests. 
\par One promising approach to avoid category scales is to present a pair of stimuli (usually side by side) and require every subject to select the image which has the highest image quality. This is a rather intuitive and simple approach which can lead to more accurate and reliable results \cite{mantiuk2012} compared to category rating subjective assessment tests. The aforementioned approach is referred to as pairwise comparison (PC) and is becoming a rather popular approach for quality assessment of visual media (images, videos, and 3D models). Moreover, subjects in pairwise comparison subjective assessment tests are able to distinguish small differences in quality and are more robust to changes in viewing conditions.

\par However, a PC test usually comes with a high price of having large numbers of pairs (and thus a large duration). Actually, this number grows quadratically with the number of image stimuli which often results in a long and expensive subjective test. Consider an original/reference image that has 15 degraded versions, a complete PC test, where every stimulus is paired with any other stimulus, requires to have $\frac{n(n-1)}{2} = 105$ pairs, where $n$ is the number of stimuli (or degraded images). Since in a typical subjective test, one original/reference image is not enough, the duration of PC subjective test is rather long. This contradicts ITU recommendations, which specify that a subjective test should not exceed more than 30 minutes, otherwise, it may cause fatigue, yielding in random decision of the subjects \cite{ITU-P-913}. 

\par The preference (or binary decision) of the subjects is usually aggregated and translated into a unified scale by using some preference aggregation model. It has been shown that not every pair yields equally useful data when converted to estimated scores and thus the selection (or sampling) of a subset of pairs may reduce the duration of the test while still resulting in reliable scores. One way to perform pairwise sampling is by active sampling, where the history of the previous decisions is used in a utility function to select a pair of images for the next trial (or comparison). However, the utility function must be calculated during the subjective test limiting its applicability in many scenarios, especially those which employ simple web platforms (such as those used in a crowdsourcing scenario). Moreover, the decision is made based on the previous judgments and does not consider the intrinsic characteristics of the pairs, e.g. difference in quality, and quite often assumes that subjects have to perform the test sequentially or in some form of batches with a specific duration.

\par The objective of this paper is to propose a novel machine learning-based approach that is able to either predict the subject preferences given a certain pair of stimuli (a pair of images) or to defer the decision to human subjects. \emph{defer} pairs go to a pairwise comparison subjective test where sufficient number of subjects rank image quality. On the other hand, the preference for \emph{predict} pairs is obtained from a machine learning model which attempts to estimate the human subjects' preferences. As a result, a shorter subjective test could be obtained without any modification to the typical PC subjective assessment methodology, only the \emph{a priori} selection of the most informative pairs. In a nutshell, our main contributions are as follows:
\begin{enumerate}[topsep=0pt, leftmargin=0.2in]
\item Propose a novel solution to perform pairwise sampling before the actual subjective assessment test starts, using a machine learning approach. 
\item Exploit image characteristics for the first time, namely with objective quality assessment models, to sample the most informative image pairs and to estimate subjects' preferences while still providing a very reliable assessment.
\item Significantly reduce the length of a pairwise subjective assessment test without requiring any processing during the subjective test and thus enable easy deployment in crowdsourcing scenarios or even massively parallel scenarios (e.g. subjective tests performed in different labs).
\end{enumerate} 
This type of approach is referred to as predictive sampling, providing an alternative yet efficient way to perform pairwise subjective assessment thus overcoming the major disadvantages of previous approaches. 

\par The remainder of the paper is organized as follows: section \ref{section-Related_work}, discusses the related work while section \ref{section_architecture} introduces the proposed architecture. The entire training procedure along with the labeling strategy is explained in section \ref{section_training_proc}. The performance evaluation is presented in section \ref{section_performance_evaluation} followed by conclusions in section \ref{section_conclusion}.
\vspace{-8pt}
\section{Related Work} \label{section-Related_work}
Pairwise subjective quality assessment is a powerful methodology to conduct a reliable test \cite{mantiuk2012}. However, this type of methodology brings a significant burden in terms of the duration of the test and thus it may require significant human resources (which are costly). There are many works that try to overcome this disadvantage and many of them propose intelligent methods to select stimuli for evaluation by human subjects. Several different sampling strategies have been proposed in the literature, which can be categorized as random-based, sorting-based, or active-based sampling, the last one providing much better performance, i.e. duration with respect to reliability.

\par The naive approach is to randomly select pairs for each trial (image pair shown to one subject). This approach ignores any statistical dependency on the decisions made by subjects and the data characteristics. The sampling strategy of \cite{HRRG} is based on edges available in random regular or Erdös-Rényi graphs, and then due to incomplete and imbalanced output, HodgeRank algorithm is applied to the paired comparison result to obtain a global ranking.

\par Sorting-based approaches select pairs of images that are more close to each other in terms of quality than pairs of images with different qualities. This type of approach exploits past decisions to select the pairs of images to be evaluated by the subjects and is actually very popular. In \cite{binary-tree} it was proposed a binary tree sorting method where the left and the right sub-tree are judged to be lower and higher in quality respectively and each tree node represents the stimuli to be compared (e.g. image). During the sorting, comparisons between nodes will be performed by subjects to maintain the tree sorted and balanced. Due to the use of a binary tree, more trials will be conducted between closer tree nodes than between distant nodes. Another popular sorting-based approach is the Swiss system \cite{MCL-V} \cite{TID2013}. It involves sorting a list of images during multiple iterations by comparing adjacent images in the list. The rule is to avoid comparing any two images more than once. In this approach, the initial list of images is chosen randomly which can introduce some variability in the output.

\par Nowadays, active sampling approaches are considered the most promising, which aim to select a pair (or a set of pairs) of images according to previous 
judgments during the subjective test. The selected pair(s) can be evaluated by one or more subjects and their preference is recorded and used to select additional pairs for evaluation in the next iteration. In active sampling approaches, a utility function is typically used to measure the information gain that each pair of images will contribute to the overall performance. This helps to select the most informative pairs for evaluation by the subjects, thereby reducing the number of comparisons needed to accurately estimate the underlying preferences.

In Hybrid-MST \cite{Hybrid-MST} a hybrid active sampling strategy for pairwise sampling based on Bradley-Terry (BT) model was proposed. The information gain of each pair is measured, rather than jointly by considering all the pairs at once. The information gain is based on Kullback Lieber Divergence (KLD) between the prior distribution of the scores and posterior distribution when possible preferences for each output of the trial are considered. In Hybrid-MST, a hybrid utility function was proposed that can switch between the global maximum information gain and the minimum spanning tree methods. The latter considers pairwise comparison as an undirected graph and enables to have a sequence of pairs with just one iteration. In ASAP\cite{ASAP}, information gain is based on a probability weighting function which assigns a higher probability to pairs of images that are close in quality and a lower probability to pairs that are far apart in quality. This approach reduces the computational cost by selectively measuring the information gain for pairs of images that are most likely to provide useful information. However, inferring quality scores are based on message passing and is still rather complex. Information gain measurement in \cite{HrActive} is similar to other active sampling approaches by calculating KLD between prior and posterior distribution of the quality scores, however, the scores are inferred based on Hodge decomposition to further evaluate any inconsistency in pairwise comparison data. 

Some other studies boost PC with data obtained from a single (or double) stimulus subjective test to avoid doing a complete pairwise comparison test, selecting more informative pairs and thus obtaining finer discrimination. In \cite{ling2020strategy} a fusion strategy is introduced to combine ACR and PC subjective tests by initializing PC with ACR subjective test results to achieve higher accuracy. Furthermore, \cite{ye2014active} introduces a hybrid approach capable of merging MOS and PC tests through a probabilistic model and active sampling approach. They have shown that this type of strategy can effectively reduce the number of trials required for achieving a target accuracy.

The reliability of the subjects' decisions may also affect the utility function calculation. The work of \cite{Crowd-BT} extends the BT model to eliminate the impact of unreliable subjects. In \cite{fan2017active} it was observed that the human visual system is not able to distinguish subtle differences between two images of very similar quality. Thus, no matter how many times the pair is selected, subjects may not have the correct preference. This highlights the importance of considering the reliability of preference when designing active sampling strategies for pairwise comparison tests and thus adjusting the selection of pairs according to their ambiguity.

The speed and reliability of human subjects can affect the efficiency of the active sampling process. Moreover, the computational cost of the utility function remains a challenge, particularly in crowd-sourcing scenarios where the server may need to run the active sampling method after every comparison by subjects, being difficult to integrate into existing web frameworks. The coordination of multiple instances of the active sampling algorithm in laboratory environments can also be challenging. This is the main motivation to design a novel sampling approach that is performed before the subjective test. 
\vspace{-8pt}
\section{PS-PC Architecture}\label{section_architecture}

A straightforward but lengthy approach in a PC subjective test is to evaluate every possible pair combination of stimuli (complete test). The number of pairs increases quadratically, $O(n^2)$, with respect to the number of stimuli (degraded images) resulting in a long and expensive subjective test. The binary decision of each comparison/trial is recorded into a matrix, typically called $PCM$ (PC matrix) which is then converted to scores using some preference aggregation model. However, it is widely known that not all pairs have the same importance and some could even be discarded. This paper proposes a framework that performs pairwise sampling (or selection) based on two aspects: which are the best pairs to be selected for human evaluation and what is the human preference considering the underlying quality of each stimulus of the pair.

\par
\begin{figure*}[htbp]
\captionsetup{skip=8pt}
  \centering
  \includegraphics[scale=0.44]{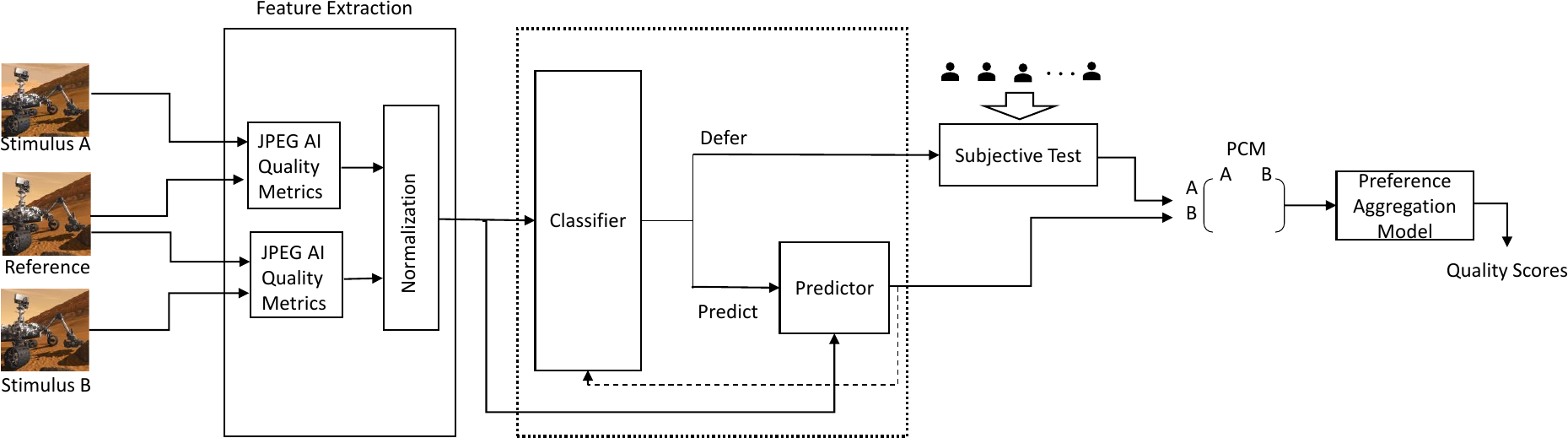}
  \caption{The Predictive Sampling Pairwise Comparison (PS-PC) framework.}
  \Description{}
  \label{Fig_framework}
  \vspace{-8pt}
\end{figure*}

 The proposed predictive sampling framework, codenamed PS-PC \footnote{Available online at \href{https://github.com/shimamohammadi/PS-PC}{https://github.com/shimamohammadi/PS-PC}}, is shown in Fig. \ref{Fig_framework} has three key parts namely, feature extraction to obtain relevant measures of quality, classifier to select some pairs for the subjective test, and a predictor that estimates the preference between the two stimuli of the pair. For every possible pair combination, the proposed framework will be run for the two stimuli of the pair. The classifier decides whether the pair should go to a subjective test or to the predictor. The predictor is trained to learn the probability of preference of the selected pair based on the features of both stimuli. The classifier decision can be interpreted as \emph{defer} or \emph{predict}. In the end, when all pairs are considered, a complete $PCM$ is obtained which can be used by the preference aggregation model to obtain quality scores. In Fig. \ref{Fig_framework} the dotted part requires a training procedure that is described in the next section.

The proposed PS-PC framework is explained, in summary, as follows:
\begin{itemize}[topsep=0pt, leftmargin=0.2in]
\item{Feature extraction and normalization}: \label{section-FE} The features used by both classifier and predictor are obtained from seven full reference image quality assessment (FR-IQA) metrics and presented in summary in table \ref{tbl_qmetrics}. These features represent the perceptual quality between the reference image and the two stimuli of the pair. The JPEG AI quality metrics, which were selected based on a rigorous evaluation done for several image compression and processing solutions (both conventional and deep learning) allow capturing the full range of properties associated with the stimuli under evaluation, not only for image compression but also for image super-resolution and denoising \cite{MichelaQoMEX2021}. These quality scores have to be first normalized according to (\ref{equ_MinMax1}) to be in the same scale \cite{scikit-learn}.
\vspace{-4pt}
\begin{equation}
x_{scaled} = \frac{x_{k}-x^{min}_{k}}{x^{max}_{k} - x^{min}_{k}}
\label{equ_MinMax1}
\end{equation}
where $x_k$ stands for the $k_{th}$ feature, $x^{min}_{k}$ and $x^{max}_{k}$ are the minimum and the maximum value in the $k_{th}$ feature respectively.

\item{Classifier}: The classifier receives as input the set of features extracted according to the feature extraction module, for each stimulus of a pair. The objective of the classifier is to perform a binary decision which is to classify as \emph{defer}: pair must be evaluated by subjects, or as \emph{predict}: subjects preference is obtained automatically without human intervention. The decision depends on the extracted features for both stimuli of the pair, the selected classifier model, and the ground-truth data. Therefore, for the training of the classifier, the ground-truth labels of \emph{defer} or \emph{predict} for each pair are required. The procedure to obtain labels for the ground-truth data is detailed in section \ref{section-lbl}. 

\par Two types of classifiers were selected, Support Vector Machine (SVM) \cite{SVM} and Extreme Gradient Boosting (XGboost) \cite{XGBoost}. Deep neural network architectures were not selected since the amount of training data available was not high, which is typically required for this type of solution. In summary, the two classifiers are presented next:
\begin{itemize}[topsep=0pt, leftmargin=0.2in]
    \item SVM: SVM finds a hyperplane that separates the data into different classes. The hyperplane is selected to maximize the margin, which is the distance between the hyperplane and the closest data points from each class. SVM uses a technique called kernel trick, where the data is transformed into a higher-dimensional space to make it linearly separable.
    \item XGBoost: XGBoost is an implementation of gradient boosting decision trees, a machine learning algorithm that works by iteratively adding models which are combined to create an overall model with higher accuracy. XGBoost uses a combination of regularized learning objectives, parallel processing, and tree pruning techniques to prevent overfitting and improve efficiency.
\end{itemize}

\item{Predictor}: The predictor is responsible to estimate the probability of preferring one stimulus over the other in a pair and is only used when a pair is classified as \emph{predict}. The output of the predictor is also used during the training process to enable the classifier to perform better decisions. This is explained in more detail in section \ref{section-lbl}. For the predictor, support vector regression (SVR) \cite{SVR} was used and trained to learn the underlying relationship between the extracted features and the probability of preference from the ground-truth data. SVR can handle both linear and non-linear regression problems and is based on the same principles as the SVM algorithm used for classification. 

\begin{table*}
\captionsetup{skip=0pt}
  \caption{JPEG AI image quality metrics}
  \label{tab:freq}
  \small
  \scalebox{0.95}{
  \begin{tabular}{p{2cm}p{16cm}} 
    \toprule
    Quality metric & Description\\
    \midrule
    IW-SSIM \cite{wang2010information} & Extension of SSIM based on information content weighted pooling, where weights are derived from statistical models of natural images.\\
    MS-SSIM \cite{MSSSIM2003} & Boosts SSIM metric by considering the variations in image resolution and viewing conditions. \\
    FSIM \cite{FSIM2011} & Exploits phase congruency and gradient information to account for local structure and contrast information.\\
    PSNR-HVS \cite{PSNRHVSM2007} & Uses DCT basis functions and calculates the maximum distortion that is not visible due to between-coefficient masking. \\
    VIF \cite{VIF2004} & Distortion measure in the wavelet domain and related to the Shannon mutual information between the degraded and original pristine image. \\
    VMAF \cite{VMAF} & Computes the quality score of several quality assessment algorithms which are fused together with an SVM algorithm. \\
    NLPD \cite{NLPD2016} & Uses a Laplacian pyramid decomposition considering different two aspects: local luminance subtraction and local contrast gain control. \\
  \bottomrule
\end{tabular}
}
\label{tbl_qmetrics}
\end{table*}
\end{itemize}
\vspace{-8pt}
\section{Training Procedure} \label{section_training_proc}
The training details of the proposed framework are described in the following sections including the labeling process to obtain ground-truth data and the classifier and predictor training. 
\vspace{-8pt}
\subsection{Ground-truth Data Creation}  \label{section-lbl}
To perform the training of the classifier, a dataset with features extracted for each stimulus and the correct labeling of each possible pair with \textit{defer} or \textit{predict} is needed. However, these labels are not available and thus a novel procedure to obtain the correct labeling from the results of a pairwise comparison test is proposed. In the next section, the preference aggregation model is presented which is important to understand the labelling procedure.

\subsubsection{Preference Aggregation Model} \label{section-Preference-aggregation}
The preference aggregation model infers scores from the subject's preferences of an already available pairwise comparison subjective assessment test. Let's assume that there is a PC test with $n$ stimuli where preferences between pairs of stimuli are arranged in a PC matrix, $PCM$. In the $PCM$, the diagonal entry is zero since a stimulus cannot be compared to itself, and the other entries correspond to the probability of preferences as follows. Let $c_{ij}$ indicate number of times stimuli $i$ is preferred over stimuli $j$. As a result, $PCM_{ij}$ corresponds to (\ref{equ_1}).

\begin{equation}
PCM_{ij}= \frac {c_{ij}}{c_{ij} + c_{ji}}
\label{equ_1}
\end{equation}

\par To infer scores from the $PCM$, several models were proposed including Bradley-Terry model \cite{BT_model}, Thurstone-Mosteller (TM) \cite{TM_model}, Borda count \cite{Borda_count}, and many others. In this paper, the BT model is utilized. According to the BT model, the probability of stimulus $i$ preferred over stimulus $j$ denoted as $Pr(i>j)$ is measured as (\ref{equ_BT}).
\begin{equation}
Pr(i > j ) = \pi_{ij} =  \frac {\pi_{i}}{\pi_{i} + \pi_{j}},\quad\text{s.t.}\quad \sum_{i=1}^{k} \pi_{i} = 1, \quad \pi_{i} >= 0
\label{equ_BT}
\end{equation}
Where $\pi_{i}$ is the score of stimulus $i$. Moreover, the score could also be expressed as $\hat{s}_i=log(\pi_i)$. Thus, $Pr(i>j)$ could be rewritten as in (\ref{equ_score}).
\begin{equation}
\pi_{ij} =  \frac{1}{1 + e^{-(\hat{s}_i - \hat{s}_j)}}
\label{equ_score}
\end{equation}
To estimate each $\hat{s}_{i}$ from the BT model,  maximum likelihood estimation (MLE) is used. The likelihood is defined as in (\ref{equ_BT2}).
\begin{equation}
L(\hat{s}|PCM) = \prod_{i<j}(\pi_{ij}) ^ {PCM_{ij}} (1-\pi_{ij})^ {PCM_{ji}}
\label{equ_BT2}
\end{equation}
The estimated scores, $\hat{s} = (\hat{s_1}, \hat{s_2}, ..., \hat{s_k})$, follow a multivariate Gaussian distribution, and the covariance matrix $\hat{\Sigma}$ is estimated using the Hessian matrix \cite{Hessian}. Therefore, the standard deviations are obtained from the covariance matrix. The output of the preference aggregation model is scores $\hat{s}$, and standard deviations $\hat{\sigma}$.



\subsubsection{Labelling}\label{section_labeling}
In the labeling process, all possible pairs are labeled as \emph{predict} or \emph{defer}. The labeling process requires a dataset where preferences between all pairs are available and thus obtained from a complete pairwise comparison test.

\begin{figure}[htbp]
 \captionsetup{skip=0pt}
  \centering
  \includegraphics[width=\linewidth]{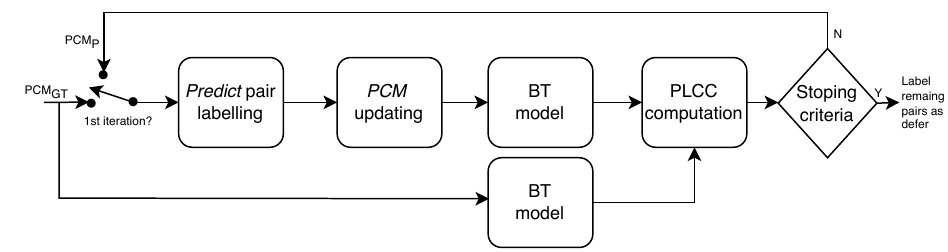}
  \caption{Main steps of the labelling algorithm.}
  \Description{}
  \label{Fig_lbl}
  \vspace{-8pt}
\end{figure}

The proposed labeling approach, shown in Fig. \ref{Fig_lbl}, involves selecting a \textit{predict} pair and removing the corresponding preferences from the $PCM$. This process continues with another \textit{predict} pair until a stopping point is reached, and then all the remaining pairs are labeled as \emph{defer}. The stopping point is defined such that only a certain reduction in the Pearson Linear Correlation Coefficient (PLCC) between the inferred scores from the ground-truth data (pristine $PCM_{GT}$) and the $PCM_P$ with some selected pairs removed is acceptable. The $\eta$ parameter defines the target PLCC that should be achieved in the labeling procedure and ranges between 0.97 and 1. Naturally, an $\eta$ value close to 1, requires labeling a significant number of pairs as \emph{defer} while a value close to 0.97 much less number of pairs (approx. 10\%), thus obtaining a much shorter test duration.

\par In general, there are two ways to remove a pair; setting its value to a constant (e.g., 0.5), to avoid numerical inconsistency of an incomplete Bradley-Terry model, or using the corresponding value from the predictor's output. In this case, the predictor's output is used whenever a pair is selected as \emph{predict}. However, for the purposes of evaluating labeling approaches, the first option is used (see section \ref{section_labeling_evaluation}).

The three different methods for labeling are:
\begin{itemize}[topsep=0pt, leftmargin=0.2in]
\item {Random-based}: The naive approach is to simply select \emph{predict} pairs assuming equal probability of all pairs. This random selection is not the best choice but it may serve as a benchmark. To ensure that the results are not biased, 50 repetitions are performed.
\item{Entropy-based}: Another selection approach is to use entropy, which is a way to measure the amount of uncertainty about an event. The idea is to calculate the entropy of each pair as in (\ref{equ_entropy}) based on the subject preferences and then, select the \emph{predict} pairs as those that have maximum entropy. In this context, entropy is the average amount of information contained in the pair and maximum entropy is used since it refers to the pairs that subjects were most uncertain about the correct answer, indicating that increasing the number of trials for these pairs may not provide much additional information for improving accuracy.
\begin{equation}
H_{ij} = -p_{ij}\log{p_{ij}} - (1-p_{ij})\log({1-p_{ij}})
\label{equ_entropy}
\end{equation}
where $p_{ij}$ refers to the probability of preferring $i$ to $j$.

\item{KLD-based}: The most principled approach towards labeling is based on the KLD, where the divergence between two probability distributions of scores is measured. The idea is to measure the divergence between the prior distribution and posterior distribution, i.e. after a pair is removed. The prior distribution represents the model's output probabilities before any pair is removed (obtained from the pristine $PCM_{GT}$), while the posterior distribution represents the model's output probabilities after a pair is removed in some iteration of the labeling algorithm (obtained from $PCM_P$). By measuring the KLD between these two distributions, it is possible to identify the pairs that are most informative for improving the model's performance. This is because pairs with a low KLD are those that are likely to have the lowest impact on the model's output probabilities and by removing the lowest KLD value (and labeling the pair as \emph{predict}) the model performance can be continuously improved. Therefore, at each labeling algorithm iteration, the KLD is computed for all possible pairs not previously labeled, to find the minimum KLD value. The inferred scores $\hat{s}$ of the BT model (see (\ref{equ_BT2})) follow a multivariate Gaussian distribution and hence, KLD must be measured between the two multivariate Gaussian prior and posterior distributions. Since the covariance matrix $\hat{\Sigma}$ is singular and thus it is not invertible, the following approximation of KLD is used \cite{ASAP}. 
\begin{equation}
\begin{array}{l}
\tilde{KLD}(\pi_{gt}, \hat{\sigma}_{gt}, \pi_{p}, \hat{\sigma}_{p}) = \\ 
\sum_{i=1}^{n}\log\frac{\hat{\sigma}_{p}}{\hat{\sigma}_{gt}} - d +\sum_{i=1}^{n}\frac{\hat{\sigma}_{gt}}{\hat{\sigma}_{p}} + \frac{1}{\hat{\sigma}_{p}}(\pi_{gt} - \pi_{p})^2
\end{array}
\label{equ_kld}
\end{equation}
In (\ref{equ_kld}) $\pi_{gth}$ and $\hat{\sigma}_{gt}$ are the ground-truth probabilities (obtained with (\ref{equ_score})), and standard deviations, respectively and represent the prior distribution. Also, $\pi_{p}$ and $\hat{\sigma}_{p}$ are calculated in the same way but for each possible pair and represent the posterior distribution. 
\end{itemize}
\vspace{-4pt}
\subsection{Classifier Training}\label{class_training}

The classifier learns to map the features to labels, i.e. the classifier receives a vector of input variables or features $f_k$, and the corresponding category output target or label $y_i$, and learns the mapping between the features $f_k$, and the label $y_i$, for each training pair $i$ (two degraded images). However, for the training of the classifier to be successful, the training set should be balanced. This doesn't happen with the proposed framework since there are much fewer \emph{defer} labels when compared to \emph{predict} labels. To address this class imbalance in the dataset, random oversampling is used on the training set. This technique involves randomly duplicating examples from the minority class until the number of examples in that class is similar to the majority class \cite{imbalanced}.

\par The SVM or XGBoost classifier hyperparameters can also significantly influence the classifier performance and must be found before training. The SVM hyperparameters considered are:
\begin{itemize}[topsep=0pt, leftmargin=0.2in]
    \item Penalty $C$: This is the regularization parameter that controls the model's ability to generalize to new data. A smaller value of $C$ will result in a larger margin hyperplane but with more misclassifications of the training data, while a larger value of $C$ will result in a smaller margin but fewer misclassifications of the training data. 
    \item Kernel: SVM can use different types of kernels to transform the input data into a higher-dimensional space where it is easier to separate the classes. Some of the popular kernels include linear, polynomial, and radial basis function (RBF). The gamma hyperparameter is specific to the RBF kernel and controls the width of the kernel. 
\end{itemize}
The hyperparameters considered in the XGBoost classifier are:
\begin{itemize}[topsep=0pt, leftmargin=0.2in]
    \item Max depth: This hyperparameter controls the maximum depth of a tree in the ensemble. A larger value will allow to obtain a more complex model well suited to the training set, but may also lead to overfitting.    
    \item Learning rate: This hyperparameter controls the step size at each iteration while moving towards the minimum of the loss function. A smaller learning rate will lead to slower but more precise convergence, while a larger learning rate will result in faster convergence but may miss the minimum. 
    \item Gamma: This hyperparameter specifies the minimum loss reduction required to make a partition on a leaf node of a tree (split).
    \item Regularization lambda: This hyperparameter controls the L2 regularization term strength, the penalty term of the loss function used during model training. The model will be more conservative (with more generalization capabilities) by increasing this parameter value. 
    \item Scale position weight: This hyperparameter is used to adjust the balance of positive and negative class weights (\emph{predict} and \emph{defer}) in this binary classification problem. 
\end{itemize}
The hyperparameters of the classifier are found using grid search which is one of the most popular techniques, where a range of values (interval) was defined for each hyperparameter. For the SVM classifier, the penalty and gamma interval was 0.05 to 0.5 and 0.01 to 1, respectively. The RBF kernel was always used. For the XGBoost, the intervals are 1 to 4 for max depth, 0.05 to 0.1 for learning rate, 0.01 to 1 for gamma, and 1 to 10 for regularized gamma. The scale position weight parameter was kept fixed according to\cite{XGBoost}: $\frac{\sum(\text{\emph{Defer} instances})}{\sum(\text{\emph{Predict} instances})}$. All possible combinations of the hyperparameters inside each interval were evaluated to select the combination that has the highest accuracy. Regarding the SVM classifier, the F1 score (harmonic mean of the precision and recall) was used to measure accuracy, while for the XGBoost classifier the area under the ROC (Receiver Operating Characteristic) curve was used.  
\vspace{-8pt}
\subsection{Predictor Training}
The predictor is responsible to estimate the probability of preference for \emph{predict} labeled pairs with the SVR regression model. During SVR training, the objective is to find a regression function that accurately estimates the probability of preference using the previously computed features (see section \ref{section-FE}). The dependent variable is the probability of preference from the ground-truth data. For each training pair $i$ (two degraded images), SVR receives a vector of input variables (features) $f_k$ of each stimulus and the corresponding scalar output target (probability of preference) $y_i$, and attempts to learn a function which maps $f_k$ to $y_i$. In SVR, a kernel can also be used to map the input data into a high dimensional feature space where data is more linearly correlated with the outputs. In this work, the widely popular RBF kernel was used. 
\vspace{-8pt}
\section{Performance Evaluation} \label{section_performance_evaluation}

In this section, the proposed PS-PC framework is thoroughly evaluated, every component individually but also jointly regarding state-of-the-art. First, the several labeling approaches (see section \ref{section_labeling}) and classifier models (see section \ref{class_training}) are evaluated. Then, the overall PS-PC solution (with the best performing labeling and classifier methods) performance is compared to state-of-the-art algorithms. A Cross-dataset evaluation was also performed to assess the PS-PC generalization capabilities along with an ablation study.
\vspace{-4pt}
\subsection{Test Conditions} \label{subsection_test_conditions}
This section describes the test conditions, namely the datasets and the procedure used to evaluate the proposed PS-PC framework.

\subsubsection{Datasets}\label{datasets}
A publicly available dataset called Pairwise Comparison Image Quality Assessment (PC-IQA) \footnote{Available \href{https://github.com/qianqianxu010/KDD2017/tree/master/dataset}{here}} was selected \cite{Xu2012}. The PC-IQA dataset was obtained with a crowd-sourcing pairwise subjective test using images obtained from the LIVE \cite{Live} and IVC \cite{Callet2005} datasets. In total, 15 reference images and 15 distorted versions of each reference are present in the dataset, with distortions including JPEG2000, JPEG, White Noise, Gaussian Blur, Fast Fading Rayleigh, Locally Adaptive Resolution (LAR) Coding, and Blurring. In summary, for each reference, $\frac{n(n-1)}{2} = 120$ pairs are available which sums to 1800 for the complete dataset. This dataset is complete (all pairs are compared) but imbalanced, which means that each pair was evaluated with a different number of trials (actual comparisons by subjects).

\par For the cross-dataset evaluation, two different datasets obtained using pairwise comparison methodology were used, the TID2013 \cite{TID2013} and the PieAPP \cite{PieAPP} test set of 2018. These datasets were not used for training. The TID2013 dataset consists of 25 references, each with 120 degradations and Swiss design was employed to reduce the number of trials (but all pairs are compared at least nine times). The test set of the PieAPP dataset comprises 40 reference images and 15 degraded versions randomly selected from 31 different distortion types. No pairwise sampling method was used and thus it's a complete and balanced dataset. For the PieAPP dataset, the probability of preference between two stimuli of a pair is available. However, for the TID2013 dataset, only quality scores ranging from 0 to 9 are available. To address this limitation, the probability of preference between each pair of images was calculated with (\ref{equ_score}), which is valid since scores were obtained from a pairwise comparison subjective test. Note also that a complete pairwise comparison test on the TID2013 and PieAPP datasets requires a total of 178,500 pairs and 4,800 pairwise comparisons, respectively which is impossible to realize in practice without employing a pairwise sampling method. 

\subsubsection{Evaluation Procedure} 
The evaluation procedure is rather straightforward, after the PS-PC algorithm performs the selection of pairs, the probability of preference for the selected \emph{defer} pairs was obtained from the ground-truth data, while for the \emph{predict} pairs was obtained with the proposed predictor. This avoids performing a subjective assessment test. 

\par The evaluation of the PS-PC classifier (in section \ref{class_evaluation}), the comparison with state-of-the-art (in section \ref{state_of_the_art_comparison}) and the ablation study (in section \ref{ablation_study}) was performed using cross-validation to ensure that the model is tested on data that it has not seen during training and thus the performance evaluation is unbiased and representative. The PC-IQA dataset is split into 5 non-overlapping folds, where each fold is used as a test set while the rest of the data is used as the training set. The result is predicted quality scores for all stimuli contained in the testing fold. Each fold contains three references (with all the associated degraded images, i.e. stimulus) and naturally, this process is repeated 5 times, with each one of the 5 folds used exactly once as test data, thus including all 15 reference images.
\vspace{-8pt}
\subsection{Experimental Results}
The performance metrics for all experiments in this section are PLCC and Spearman's rank correlation coefficient (SROCC). The correlation is calculated between the ground-truth scores (obtained from the preference aggregation model) and the estimated scores of the proposed PS-PC solution unless stated otherwise.  

\subsubsection{Labeling Evaluation}\label{section_labeling_evaluation}
The labeling algorithms presented in section \ref{section_labeling} are evaluated on the entire PC-IQA dataset without any stopping criteria to better understand the performance for a wide range of \emph{predict} pairs numbers and thus subjective test length. Therefore, the PLCC performance is reported every time that a pair is removed from the $PCM$ and labeled as \emph{predict} according to the labeling algorithm. To avoid any bias for the random-based labeling approach, the average results of 50 iterations were used.

\par
\begin{figure}[!htbp]
\captionsetup{skip=0pt}
\centerline{  
    \includegraphics[scale=0.38]{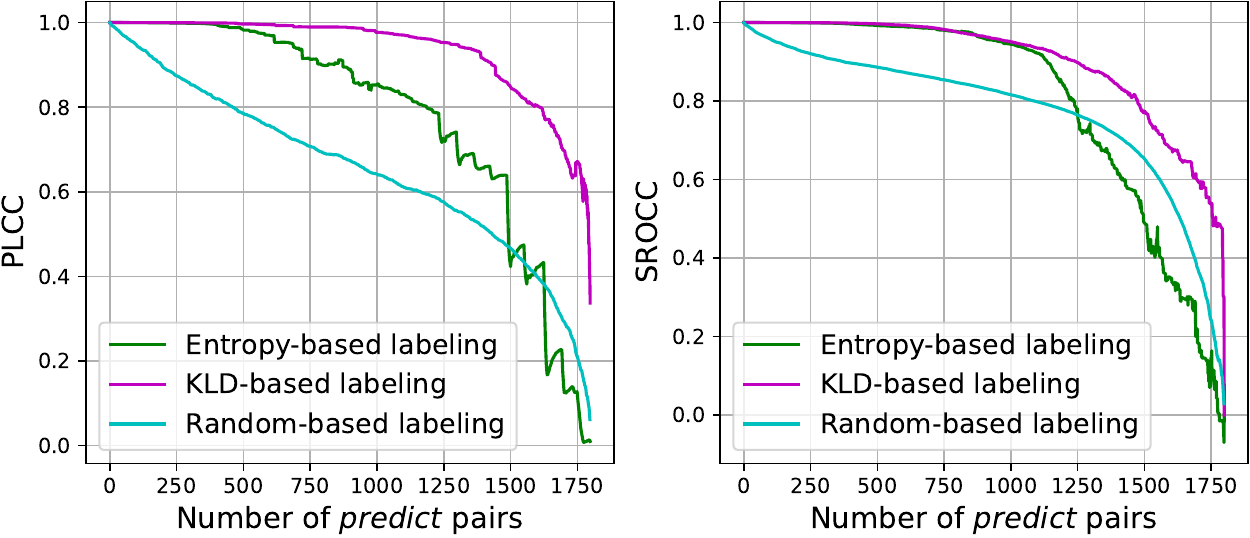}
}
\caption{Labeling algorithm evaluation.}
\label{fig_compare_lbl}
\vspace{-12pt}
\end{figure}

 Fig. \ref{fig_compare_lbl} compares the three labeling approaches in terms of PLCC and SROCC. The figures show that KLD-based labeling provides the best selection of pairs since PLCC gradually reduces and is always above the other labeling approaches. The experimental results also suggest that, initially, entropy-based labeling has a higher correlation than random-based labeling, but as more pairs are removed, the correlation is worse than the average random-based labeling. 
\vspace{-4pt}

\subsubsection{Classifier Evaluation} \label{class_evaluation}
To evaluate the two alternatives for the classifier, SVM and XGBoost, five different classifier models were trained. In each model, the ground-truth labels were obtained by having different $ \eta = \{0.97, 0.98, 0.985, 0.99, 0.995\}$ in the labeling algorithm which defines the tradeoff between the number of pairs selected as \emph{defer} (and thus the subjective test length) and the accuracy of the scores in the subjective test. An ideal pairwise sampling framework is expected to have a performance equal to $\eta$.

\par
\begin{figure}[ht]
 \captionsetup{skip=0pt}
  \centering
  \includegraphics[scale=0.38]{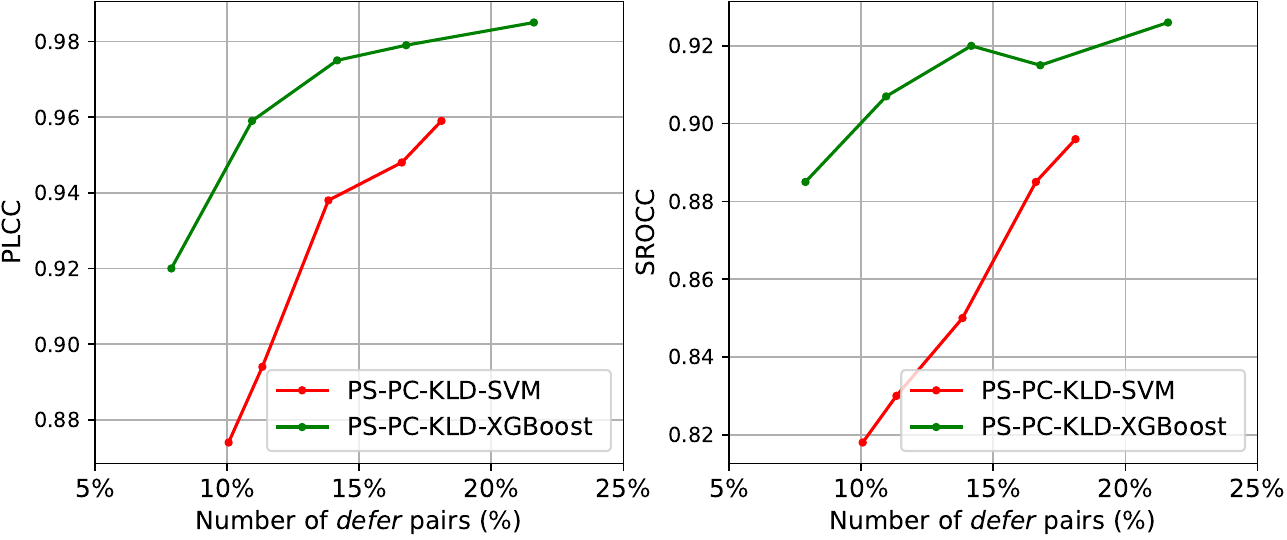}
  \caption{Classifier models evaluation.}
  \label{Fig_clf_experiment}
  \vspace{-12pt}
\end{figure}

 The experimental results are shown in Fig. \ref{Fig_clf_experiment}. In this figure, the horizontal axis is the percentage of the number of \emph{defer} pairs (that go to the subjective test) each classifier selects for the complete PC-IQA dataset (total of 1800 pairs). The experimental results show that XGBoost is the best choice. 
\vspace{-4pt}

\subsubsection{State-of-the-art Evaluation}\label{state_of_the_art_comparison}
The PS-PC performance evaluation was compared to several state-of-the-art algorithms: \textbf{Hybrid-MST} \cite{Hybrid-MST}, \textbf{HR-Active} \cite{HrActive}, \textbf{ASAP} \cite{ASAP}, \textbf{Crowd-BT} \cite{Crowd-BT}, and \textbf{Swiss-design} \cite{TID2013} are the selected benchmarks. The first three are active sampling methods and the last is a sorting-based method which is often used in subjective image quality assessment studies \cite{Men2021SubjectiveIQ, PCS-paper, TID2013, MCL-V}.

The PS-PC framework uses the probability of preference for \emph{defer} pairs obtained from the ground-truth data (the result of multiple trials in a subjective test) whereas the state-of-the-art methods only perform a single trial for some selected pairs in each iteration of the algorithm. Therefore, the comparison should be based on the number of trials instead of the number of \emph{defer} pairs. To enable a fair comparison, the number of trials for the \emph{defer} pairs selected by the PS-PC framework is recorded and summed for each $\eta$. This will be the budget of trials for the benchmark pairwise selection methods. It is important to note that benchmarks use the raw decisions of the ground-truth data (randomly selecting a subject comparison) and not the probability of preference. Since these raw decisions are obtained randomly, an average of 50 iterations is used.

\begin{figure}[ht]
\captionsetup{skip=0pt}
  \centering
  \includegraphics[scale=0.38]{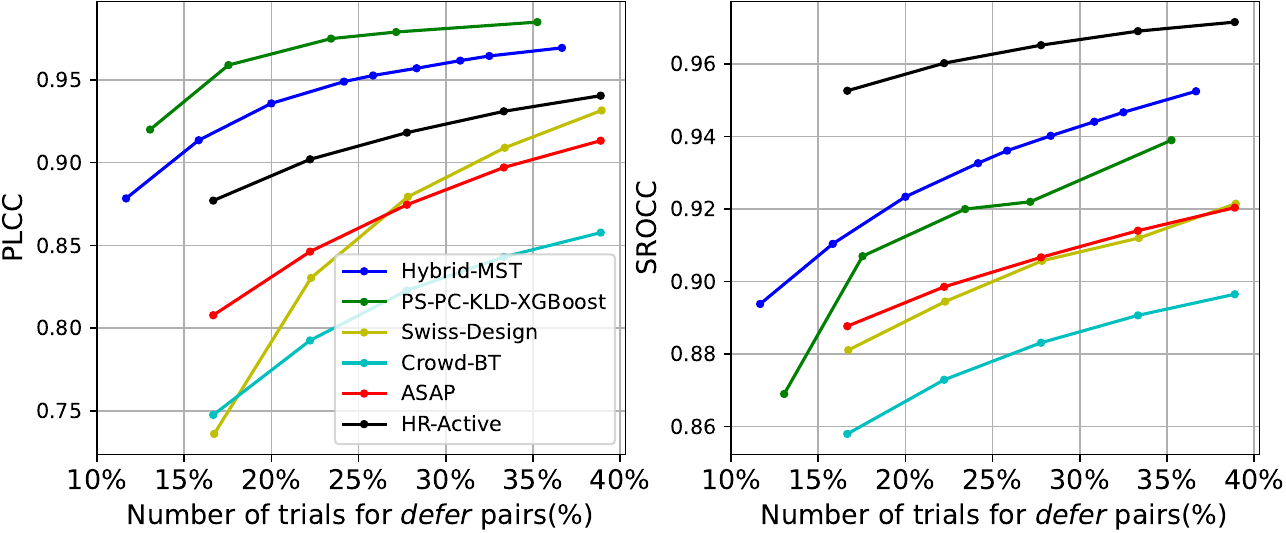}
  \Description{}
  \caption{PS-PC performance against selected benchmarks.}
  \label{Fig_SOA}
  \vspace{-12pt}
\end{figure}

In Fig. \ref{Fig_SOA}, the horizontal axes is the number of trials in percentage where the maximum number of trials is set to $\frac{n(n-1)}{2}\times15$ with 15 being the number of subjects. As shown in the results, for the PLCC correlation measure, PS-PC is the best choice followed by Hybrid-MST and both have far better correlations than Swiss-Design and Crowd-BT. However, if SROCC is considered as the correlation metric, HR-Active followed by Hybrid-MST has higher performance than PS-PC. This is actually because PS-PC was trained based on the PLCC metric and was actually expected. For the training of the PS-PC framework, other performance metrics, such as SROCC (or weighted SROCC with PLCC) could have been selected instead. It is ultimately a choice of the subjective test designer.
\vspace{-4pt}

\subsubsection{Ablation Study Evaluation}\label{ablation_study}
The PS-PC framework has both classifier and predictor modules which might have a different impact on the overall correlation performance and therefore an ablation study to further validate the performance of each proposed module was done. Moreover, a random classifier was integrated in the PS-PC framework to better understand the impact of this module. In summary, the following configurations were considered.
\begin{itemize}[topsep=0pt, leftmargin=0.2in]
\item Classifier only: To individually evaluate the performance of the classifier, the predictor in the PS-PC framework outputs always a preference probability of 0.5. This is done to evaluate the performance of the classifier module alone and thus its ability to classify the pairs.
\item Predictor only: To individually evaluate the performance of the predictor, the classifier in the PS-PC framework is not present, meaning that every pair is classified as \emph{predict}. This is done to evaluate the performance of the predictor module alone without the influence of the classifier.
\item RndClass+predict: This corresponds to a benchmark where a random classifier is used to classify the pairs, instead of the proposed classifier on top of the proposed predictor. This is done to evaluate the effectiveness of the trained classifier of the PS-PC framework.
\end{itemize}
The ablation study experimental results are shown in Table \ref{table_ablation_study} for different values of $\eta = \{0.97, 0.98, 0.985, 0.99, 0.995\}$. This ablation study shows that both classifier and predictor play an important role in the overall performance. With just the predictor, PLCC is 0.85 whereas with just the classifier goes up to 0.94, naturally at the cost of selecting a high number of pairs. Neither of these solutions is a good choice for pairwise sampling, mostly due to lack of performance or a high number of \emph{defer} pairs. Moreover, the random classifier and predictor barely improve the performance compared with the predictor only case, which shows the importance of having an accurate classifier. From these results, it is clear that the best option is the proposed PS-PC framework, which has a loss of only 0.05 for a target PLCC $\eta$ of 0.97, which corresponds to the selection of just 8\% of the pairs. This loss can be even further reduced by allowing the selection of more pairs (as much as 0.01).
\vspace{-8pt}
\begin{table}[htb]
\captionsetup{skip=0pt}
\caption{PS-PC ablation study.}
\centering
\scalebox{0.75}{
\begin{tabular}{|l|c|c|c|c|c|c|}
\hline
\multirow{2}{*}{Module} & \multirow{2}{*}{Metric} & \multicolumn{5}{c|}{Models} \\
\cline{3-7}
& & $\eta = 0.97$ & $\eta = 0.98$ & $\eta = 0.985$ & $\eta = 0.99$ & $\eta = 0.995$\\
\hline
\multirow{2}{*}{Classifier only}    &PLCC  &0.67  &0.85   &0.91   &0.93   &0.94   \\
                               &SROCC &0.44  &0.60   &0.63  &0.68    &0.75   \\                          
\hline
\multirow{2}{*}{Predictor only} &PLCC    & \multicolumn{5}{c|}{0.85}   \\
                          &SROCC    & \multicolumn{5}{c|}{0.83}  \\                          
\hline
\multirow{2}{*}{RndClass+predict} &PLCC    &0.89  &0.88   &0.89  &0.87   &0.88     \\
                               &SROCC  &0.85  &0.85   &0.85  &0.86   &0.86    \\                         
\hline
\multirow{2}{*}{PS-PC} &PLCC  &0.92  &0.96   &0.97  &0.98   &0.99    \\
                      &SROCC  &0.87  &0.90   &0.92  &0.922   &0.94    \\                         
\hline
\end{tabular}
}
\label{table_ablation_study}
\vspace{-14pt}
\end{table}
\vspace{-4pt}
\subsubsection{Cross-dataset Evaluation}
To examine the generalization of the PS-PC framework, the PieAPP and TID2013 datasets \footnote{Available at \href{https://github.com/prashnani/PerceptualImageError}{PieAPP}, and \href{https://www.ponomarenko.info/tid2013.htm}{TID2013}} (described in \ref{datasets}) are used just for testing. None of the images of these datasets were used in the training procedure, only the images of the PC-IQA dataset. In this case, cross-validation was not used and all reference and degraded images of the PC-IQA dataset represent the training set. Moreover, these datasets include very different types of degradations compared to PC-IQA, for example, TID2013 includes different types of noises, transmission errors, contrast changes, other codecs, etc. This is a challenging scenario for PS-PC which has never seen such types of errors.

\par The results are shown in Table \ref{tabel_cross_dataset}. As shown, PS-PC performance is lower for TID2013 and PieApp but not very significantly, using 62\% of the selected pairs in the TID2013 dataset results in 0.95 PLCC, while 46\% of the selected pairs in the PieAPP dataset, have a correlation of 0.89 PLCC. The performance is lower for the PieApp dataset since it includes very different types of degradations, e.g. complex artifacts from computer vision and image processing algorithms.

\begin{table}[htb]
\captionsetup{skip=0pt}
\caption{Cross dataset evaluation}
\centering
\scalebox{0.8}{
\begin{tabular}{|l|c|c|c|c|c|c|}
\hline
\multirow{2}{*}{Dataset} & \multirow{2}{*}{Metric} & \multicolumn{5}{c|}{Models} \\
\cline{3-7}
& & $\eta = 0.97$ & $\eta = 0.98$ & $\eta = 0.985$ & $\eta = 0.99$ & $\eta = 0.995$\\
\hline

\multirow{3}{*}{TID2013}    &PLCC  &0.83  &0.85   &0.90   &0.92   &0.95   \\
                            &SROCC &0.82  &0.83   &0.90  &0.93    &0.95   \\ 
                            &\emph{Defer} Pairs &7\%  &17\%   &35\%  &44\%    &62\%   \\
\hline

\multirow{3}{*}{PieAPP}        &PLCC    &0.80  &0.84   &0.86  &0.89   &0.89     \\
                               &SROCC  &0.44  &0.49   &0.53  &0.60   &0.62    \\
                               &\emph{Defer} Pairs  &9\%  &22\%   &26\%  &38\%   &46\%    \\ 
\hline

\multirow{3}{*}{PC-IQA}        &PLCC    &0.92  &0.96   &0.97  &0.98   &0.99     \\
                               &SROCC  &0.87  &0.90   &0.92  &0.922   &0.94    \\
                               &\emph{Defer} Pairs  &8\%  &11\%   &15\%  &17\%   &22\%    \\ 
\hline
\end{tabular}
}
\label{tabel_cross_dataset}
\vspace{-12pt}
\end{table}
\vspace{-4pt}
\section{Conclusions and Future Work} \label{section_conclusion}
This paper proposes a novel predictive sampling framework based on machine learning to perform pair sampling for a pairwise comparison subjective test. The objective is to select a subset of pairs without compromising its performance. To achieve this, the PS-PC framework uses JPEG AI image quality metrics to extract features, which were used as input for a classifier and predictor. The classifier determines whether some pair of images should be subjectively evaluated or not; in the latter case, a predictor computes the probability of preference between the two stimuli of the pair. The performance evaluation shows that the PS-PC framework outperforms relevant state-of-the-art and that both predictor and classifier contribute to the final performance of the proposed solution. Moreover, the proposed solution selects pairs of images for subjective assessment \emph{a priori} and does not require to be run during the subjective test and thus much more simple to deploy in crowdsourcing scenarios. As future work in the near-term, a large-scale new pairwise comparison dataset will be constructed to learn new types of distortions and thus improve the generalization capabilities of this solution. This work is seminal since many different research directions could be followed next, such as the use of deep-learning network architectures as well as the exploitation of past subject decisions in this type of framework, for example with reinforcement learning techniques.
\vspace{-14pt}
\begin{acks}
This work is funded by FCT/MCTES through national funds and when applicable co-funded EU funds under the project DARING with reference PTDC/EEI-COM/7775/2020.
\end{acks}
\bibliographystyle{unsrt}
\balance
\bibliography{reference}
\end{document}